\documentclass[%f
 reprint,
superscriptaddress,
 amsmath,amssymb,
 aps,
]{revtex4-2}

\usepackage{xcolor}
\usepackage{graphicx}% Include figure files
\usepackage{dcolumn}% Align table columns on decimal point
\usepackage{bm}% bold math

\def\Eobject{{\bf E}^{\rm{obj}}}

\def\Emeasuredx{E_x}
\def\Emeasuredxtilde{\tilde{E}_x}

\def\Ereconstructedx{E_x^{\rm SL}}

\def\kmax{k_{\rm max}}
\def\ie{{\em i.e.}}
\def\eg{{\em e.g.}}
\def\cf{{\em cf.~}}
\def\kk{{\bf k}}
\def\rr{{\bf r}}
\begin{document}

\preprint{APS/123-QED}

%\title{So close, yet so far: subwavelength imaging via evanescent wave amplification in the radiating near field}

\title{Subwavelength terahertz imaging via virtual superlensing in the radiating near field}

\author{Alessandro Tuniz}
\affiliation{%
Institute of Photonics and Optical Science, School of Physics, University of Sydney, NSW 2006, Australia
}%
\affiliation{The University of Sydney Nano Institute, The University of Sydney, NSW 2006, Australia}%
 \email{alessandro.tuniz@sydney.edu.au}
\author{Boris T. Kuhlmey}%
\affiliation{%
Institute of Photonics and Optical Science, School of Physics, University of Sydney, NSW 2006, Australia
}%

\date{\today}

\begin{abstract}
Paradoxically, imaging with resolution much below the wavelength $\lambda$ -- now common place in the visible spectrum -- remains challenging at lower frequencies, where arguably it is needed most due to the large wavelengths used.  Techniques to break the diffraction limit in microscopy  have led to many breakthroughs across sciences, but remain largely confined to the optical spectrum, where near-field coupled fluorophores operate. At lower frequencies, exponentially decaying evanescent waves must be measured directly, requiring a tip or antenna to be brought into very close vicinity to the object. This is often difficult, and can be problematic as the probe can perturb the near-field distribution itself. Here we show the information encoded in evanescent waves can be probed further than previously thought possible, and a truthful image of the near-field reconstructed through selective amplification of evanescent waves - akin to a virtual superlens reversing the evanescent decay. We quantify the trade-off between noise and measurement distance, and  experimentally demonstrate reconstruction of complex images with subwavelength features, down to a resolution of $\lambda/7$ and amplitude signal-to-noise ratios below 25dB between 0.18--1.5\,THz. Our procedure can be implemented with any near field probe far from the reactive near field region, greatly relaxes experimental requirements for subwavelength imaging in particular at sub-optical frequencies, and opens the door to non-perturbing near-field scanning.
%We present a universal method for amplifying evanescent waves in the radiating near field, reconstructing sub-wavelength images akin to a virtual superlens. We quantify the trade-off between noise and measurement distance, demonstrating complex image reconstruction with subwavelength features down to $\lambda/7$ and amplitude signal-to-noise ratios below 25 dB between 0.18--1.5 THz. Our procedure applies to any near field probe, greatly relaxes experimental requirements for subwavelength imaging at sub-optical frequencies, and opens the door to non-perturbing near-field scanning.
\end{abstract}

\maketitle

\onecolumngrid

%%%%%%%%%%%%%%%%%%%%%%%%%%%%%%%%

\section*{Introduction}

The diffraction limit is a consequence of the evanescent decay of high spatial frequencies in standard materials~\cite{novotny2012principles}. Conventional imaging techniques that collect light far from an object are typically bound by this limit, and much effort has been invested in developing ways to overcome it. 
Many techniques now provide resolutions well below the diffraction limit~\cite{schermelleh2019super, adams2016review}, relying either on near-field probing through a scanning tip%(near field optical microscopy)
~\cite{keilmann2004near, chen2019modern}, stochastic sets of scatterers or fluorophores in the immediate vicinity of the object to be imaged~\cite{zhuang2006STORM,betzig2006palm}, or nonlinear effects~\cite{frischwasser2021real, hell1994sted}, 
only really possible in the optical spectrum. Methods to reconstruct sub-diffraction details from linear far fields also exist, but typically require some prior knowledge or assumptions on the nature of the object~\cite{Gazit2009sparse}.

Lower frequency sub-wavelength imaging techniques (\eg, GHz, THz) typically rely on scanning antennas in an object's near field~\cite{belov2006subwavelength, sawallich2016photoconductive}.  
Imaging at terahertz frequencies (0.1-10\,THz) would particularly benefit from any improvement in the ability to image below the diffraction limit~\cite{lee2022frontiers}, due to its many applications in biomedicine~\cite{ lee2018ultrasensitive, yang2016biomedical, heo2020identifying, smolyanskaya2018terahertz}, which is hindered by established experimental challenges~\cite{markelz2022perspective}.
We refer the reader to Ref.~\cite{lee2022frontiers} for a recent review on recent developments in terahertz imaging techniques. 
The state-of-the art, in terms of resolution ($<$100\,nm), is given by nanoscale THz scanning probe microscopy~\cite{cocker2021nanoscale}, a sophisticated technique that can only access small-area planar surfaces.
A practical alternative which can be incorporated into any commercial THz time domain spectroscopy setup~\cite{menlo}, is provided by near-field photoconductive detector antennas~\cite{sawallich2016photoconductive, protemics}, which directly probe both amplitude and phase over ${\rm cm}^2$ areas with resolutions of order $10\,\mu {\rm m}$ at the site of the antenna. However, under standard laboratory conditions, it is common for such antennas to be hundreds of micrometers away from the object to be imaged~\cite{stefani2018terahertz, stefani2022flexible}. Such large measurement distances are arguably desirable, since local fields can be heavily perturbed by the presence of near-field probes~\cite{arango2022cloaked}. Furthemore, such antennas are expensive and delicate, making near-contact scans a risky proposition. At THz frequencies, such distances are associated with the radiating near field region (\ie, distances between $\lambda/2\pi$ and $\lambda$ from the object~\cite{yaghjian1986overview}), where high spatial frequencies decay significantly, preventing genuine sub-wavelength imaging~\cite{novotny2012principles}. 

Ideally, the exponential decay of the evanescent field would be avoided or reversed. %A refined understanding of structured optical materials led to the development of 
This can be achieved with so-called superlenses (SL)~\cite{pendry2000negative, fang2005sub} and hyperlenses (HL)~\cite{jacob2006optical}, which respectively amplify or propagate the evanescent fields, and which have been demonstrated over much of the electromagnetic spectrum~\cite{belov2006subwavelength, lemoult2012polychromatic, tuniz2013metamaterial, casse2010super, fang2005sub, liu2007far, rho2010spherical}. Such approaches still carry two challenges: (i) the resolution of the highest spatial frequencies is adversely affected by even modest losses~\cite{podolskiy2005near}; (ii) most geometries transfer the near field information without spatial magnification which converts evanescent waves into propagating waves~\cite{smolyaninov2007magnifying, tuniz2013metamaterial}. Fields must still be measured in the near field of the SL or HL~\cite{taubner2006near}, shifting the problem rather than solving it. At THz frequencies, anisotropic metamaterials have been used for sub-wavelength propagation of near-field information across finite slabs~\cite{tuniz2013metamaterial,kaltenecker2016ultrabroadband, huang2020amplifying} using Fabry-Perot resonance-induced evanescent amplification~\cite{tuniz2014imaging, tuniz2015two}, but a THz SL which truly amplifies decayed evanescent fields has so far been beyond reach. 

%%%%%%%%%%%%%%%%%%%%%%%%%%%%%%%%%%%%%%%%%%
\begin{figure}[t!]
\centering
\includegraphics[width=0.6\textwidth]{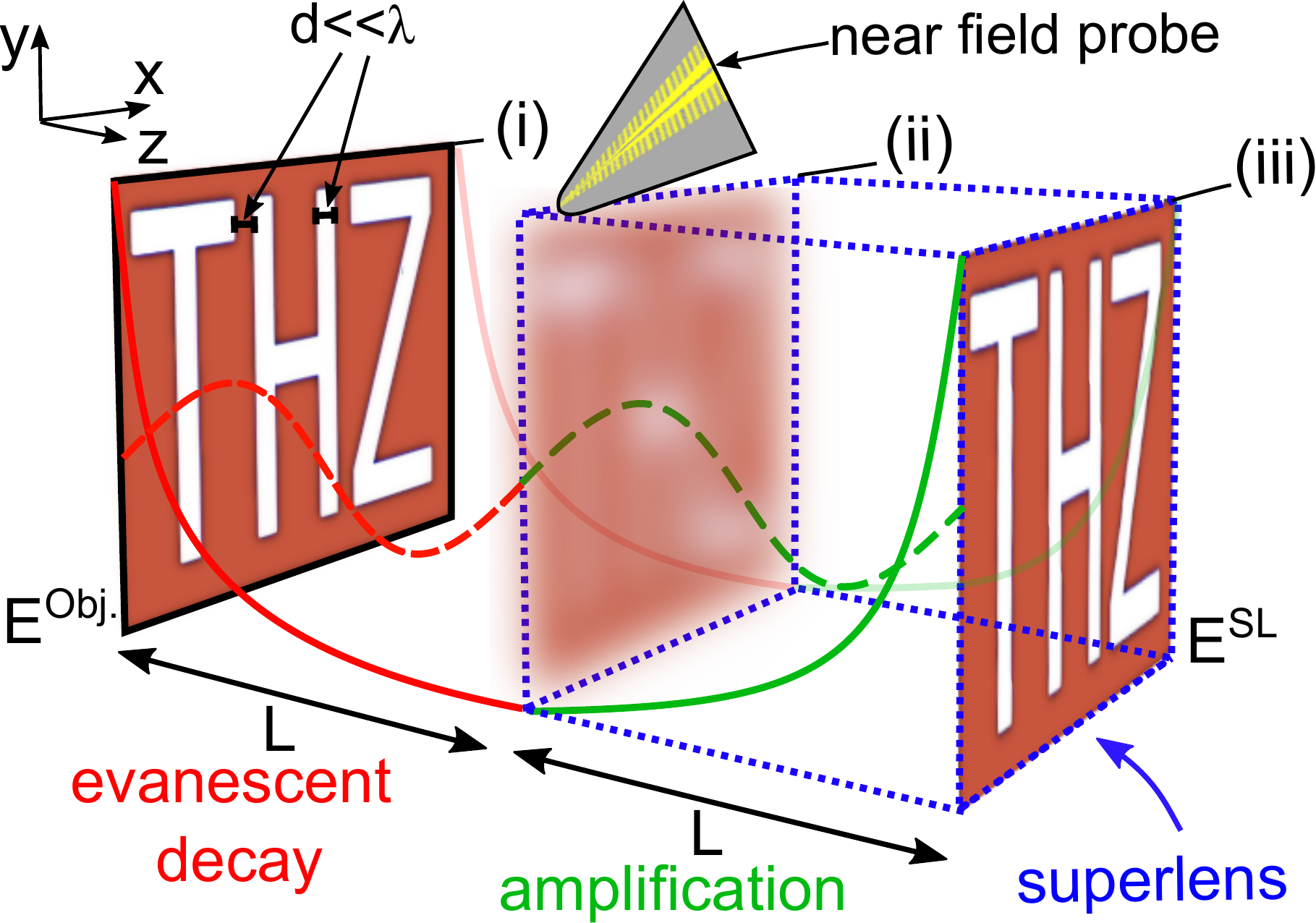}
\caption{Concept schematic of virtual superlens. (i) Sub-wavelength spatial features are carried by evanescent waves which exponentially decay over a $L$ (red). (ii) The resulting lower-resolution image is detected by a near field probe. The collected evanescent fields are then numerically amplified over  $L$ (green), leading to (iii) the original image, analogously to a superlens (blue). Wavelength-scale information is carried by propagating waves (dashed).}
\label{fig:fig1}
\end{figure}
%%%%%%%%%%%

In the absence of noise, reversing evanescent decay does not necessarily require physical devices, but can be done numerically instead. Evanescent waves contribute to the field measured at any distance, superimposing high spatial frequency fluctuations on top of low spatial frequency radiating fields. If the fluctuations can be resolved, they can also be amplified to regain the original field. To our knowledge, such a scheme has not been considered, presumably because high-spatial frequencies commonly decay well below instrument noise. Here we show that practical noise levels allow for a useful extraction of decayed information in the radiating near field region, also providing a way for increasing the resolution in the reactive near field region. In effect, we experimentally demonstrate a virtual superlens through post-processing, reconstructing previously indiscernible sub-wavelength spatial features contained within complex images, with demonstrated resolution down to $\lambda/7$. Our approach is general, provided that low-noise, phase-resolved fields can be measured. This opens the possibility of measuring near fields without perturbing them - which would be particularly useful when resolving fields in structures that are sensitive to perturbations, \eg, high-Q/topological resonances~\cite{van2021unveiling, yang2022topology} and photonic crystal defects~\cite{akiki2020high}. 

\section*{Results}

\subsection*{Approach and implementation}

Figure~\ref{fig:fig1} shows a schematic of our approach, which aims to image a planar source object possessing subwavelength features with a field $\Eobject (x,y,z=0)$. The total field at $\rr=(x,y,z)$ is given by a Fourier expansion~\cite{pendry2000negative}
\begin{equation} \label{eq:Esuperp}
{\bf{E}}(\rr)  =  \sum_{\sigma}\iint_{k_x, k_y}  {\bf \tilde{E}}^\sigma(k_x,k_y) \mathrm{exp}(i\kk\cdot\rr)dk_xdk_y,
\end{equation}
where $\sigma$ sums over polarizations, $\kk=(k_x,k_y,k_z$), 
\begin{equation}
k_{z}  =  (k_0^2 -k_x^2 - k_y^2)^{1/2},
\label{eq:kz}
\end{equation}
$\bf \tilde{E}^\sigma$ can be obtained from the Fourier transform of $\Eobject(x,y,z=0)$, and $k_0=2\pi/\lambda$ is the free space wavenumber. Propagating waves (Fig.~\ref{fig:fig1}, dashed) carry information emerging from spatial frequencies satisfying $k_x^2 + k_y^2<k_0^2$ and impose a lower limit on the spatial features $d$ which can be resolved in the far field. Evanescent waves (curves) carry sub-wavelength spatial frequencies satisfying $k_x^2 + k_y^2>k_0^2$ and exponentially decay in free space. As a result, the fine details of an image possessing spatial features $d\ll \lambda$, detected by a near field probe at a distance $z=L$, cannot be resolved. This process can be straightforwardly reversed via the transformation $(x,y,z) \rightarrow (-x,-y,-z)$ over a subsequent length $L$, by numerically reversing the phase accumulated by the propagating waves and amplifying the evanescent waves (green curves in Fig.~\ref{fig:fig1}). In practice, we measure an $x$-polarized field $\Emeasuredx(x,y,z=L)$. Starting from Fourier components of the measured field $\Emeasuredxtilde%^{\sigma}
(k_y,k_z)$, the electric field after the virtual SL is given by:

\begin{equation}%\begin{split}
\Ereconstructedx(x,y) \!= \! %\sum_{\sigma}
\iint_{k_x, k_y}  \! \! \! \! \! \! \Emeasuredxtilde
%^\sigma
(k_x,k_y)\mathrm{exp}[-i(k_x x+k_y y+k_{z} L )dk_xdk_y,
%\end{split}
\label{eq:SL}
\end{equation}
where $k_z \in \mathbb{C}$ follows Eq.~\eqref{eq:kz}, with arbitrarily high spatial frequencies. For large spatial frequencies, $k_z$ is imaginary and leads to exponential amplification $\exp{|k_z|L}$. This process is equivalent to  ``superlensing''~\cite{pendry2000negative}, achieving the same effect (Fig.~\ref{fig:fig1}, blue dotted line). Simply reversing the phase without amplifying evanescent waves, that is limiting the integrals to real $k_z$, is akin to what an ideal {\em conventional} (far field) lens achieves (``lensing''). %  

%%%%%%%%%%%%%%%%%%%%%%%%%%%%%%%%%%%%%%%%%%
\begin{figure}[t!]
\centering
\includegraphics[width=0.6\textwidth]{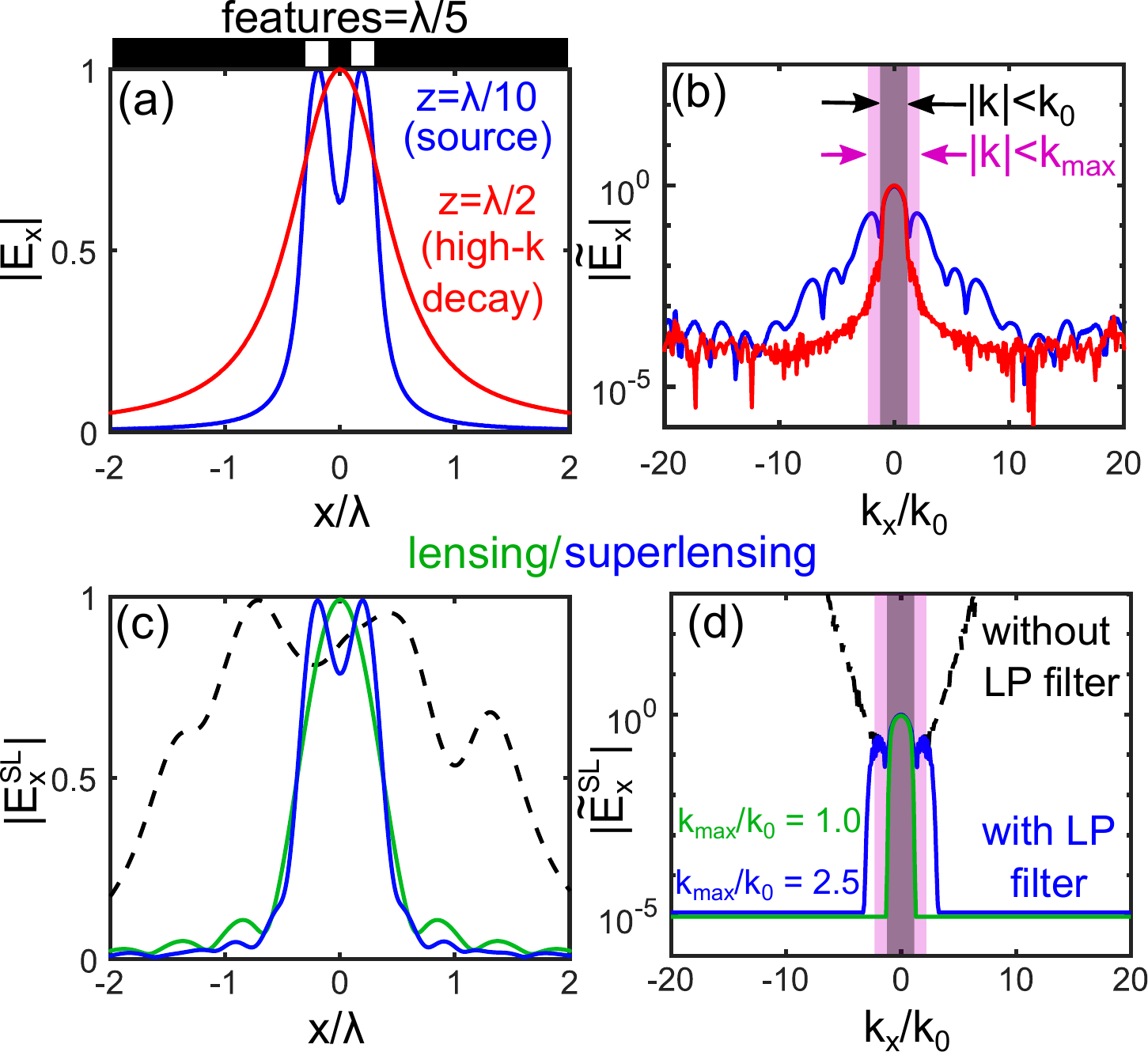}
\caption{Numerical example of virtual SL. (a) An $x$-polarized field is incident on subwavelength apertures ($d=\lambda/5$, blue), which cannot be discerned at $z=\lambda/2$ (red). (b) Associated spatial Fourier transform. Black/purple regions are propagating/evanescent. (c) Images after virtual lens ($\kmax = k_0$, green), after SL using the full spectrum (black), and using the low-passed (LP) filtered spectrum ($\kmax = 2.5 k_0$, blue). (d) Associated spatial Fourier transforms.}
\label{fig:fig2a}
\end{figure}
%%%%%%%%%%%%%%%%%%%%%%%%%%%%%%%%%%%%%%%%%%%

In  Eq.~\eqref{eq:SL} higher spatial frequencies lead to larger amplification terms so that, after amplification, high spatial frequency noise is bound to dominate over any signal at lower spatial frequencies. A simple numerical example showcases the issue: We consider 2D finite element method calculations where the domain is infinite in $y$, with TM polarized fields (non-zero magnetic field in $y$). Figure~\ref{fig:fig2a}(a) shows  $|E_x|$ emerging from two perfectly conducting apertures with width- and edge-to-edge- separation of $d=\lambda/5$. At a distance $z=\lambda/10$, the two apertures can be distinguished in the field (blue curve in Fig.~\ref{fig:fig2a}(a)). At a distance of $z=\lambda/2$ (red), however, this is no longer the case. Figure~\ref{fig:fig2a}(b) shows the associated spatial Fourier transforms. The minimum magnitude of $k_x$ required in order to resolve $d$ is given by $k_x/k_0 = \lambda/2d=2.5$. While the source's spatial Fourier spectrum extends beyond $k_{x} = 10 k_0$, at $z=\lambda/2$ (\ie, in the radiating near field region) most field components with spatial frequencies inside $|k|<\kmax$ have exponentially decayed 20\,dB below that of propagating waves, so that the apertures cannot be distinguished. 
Note that while the source spectrum (Fig.~\ref{fig:fig2a}(b), blue) is smooth, the radiating near field spectrum (Fig.~\ref{fig:fig2a}(b), red) presents numerical noise starting even at modest $k_x/k_0$ values. Using this noise level compared to the maximum signal, the amplitude signal-to-noise (SNR) is $\sim 30\,{\rm dB}$. 

We now implement the superlens procedure given by Eq.~\eqref{eq:SL} to the complex field associated with the red curves in Fig.~\ref{fig:fig2a}(a) and~\ref{fig:fig2a}(b) obtained at $z=\lambda/2$. The result is shown as a black dotted line in Fig.~\ref{fig:fig2a}(c). The image reconstruction of the two apertures has clearly failed: the associated spatial Fourier spectrum (black line in Fig.~\ref{fig:fig2a}(d)) shows that noise at high spatial frequencies has been amplified to exceed the amplitude of  any signal, including that of the propagating waves. However, not all is lost: comparing the black curve amplified spectrum with the blue curve in Fig.~\ref{fig:fig2a}(b), the amplified spectra match with the original for $k_x/k_0\lesssim 2.5$, where the signal in the evanescent spectrum was originally above the noise level (Fig.~\ref{fig:fig2a}(b), purple). We thus apply a spatial low-pass filter in $k$-space after superlensing, setting all $|k_x|>2.5k_0$ to zero. The result is shown as blue curves in Fig.~\ref{fig:fig2a}(c,d), where we find that the virtual SL now resolves the apertures as desired. In contrast, filtering all non-propagating waves as per a conventional lens, \ie, setting regions where $k>k_0$ to zero (green curves in  Fig.~\ref{fig:fig2a}(c,d)), does not allow us to resolve the apertures, even though the phase is reversed in the procedure. % -- a process we refer to as ``lensing'. 

There is thus much to be gained from amplifying evanescent waves, provided amplification is limited to spatial frequencies with signal above the noise floor. By equating the signal-to-noise ratio ${\rm SNR}$ with the amplification factor $\exp(k_z L)$ for a signal measured at a distance $L$, we obtain the maximum useful spatial frequency

\begin{equation}
\kmax  =  k_0 \sqrt{1+ \left(\frac{\lambda}{z}\frac{\log 10}{20 \pi}{\rm SNR}\right)^2},
\label{eq:kmax}
\end{equation}
where SNR (in dB) is the signal-to-noise ratio of the amplitude $|E_x|$. A detailed derivation of Eq.~\eqref{eq:kmax} is presented in the Supplementary Information, accompanied by additional numerical examples in Supplementary Fig.~1. Equation~\eqref{eq:kmax} directly estimates the maximum spatial frequency that can be retrieved, and thus the resolution that can be achieved, at any $z$ and SNR. Increasing $z$ could thus be advantageous -- reducing the near field perturbation induced by the antenna --  provided SNR increases accordingly.

%%%%%%%%%%%%%%%%%%%%%%%%%%%%%%%%%%%%%%%%%%
\begin{figure*}[b!]
\centering
 \includegraphics[width=\textwidth]{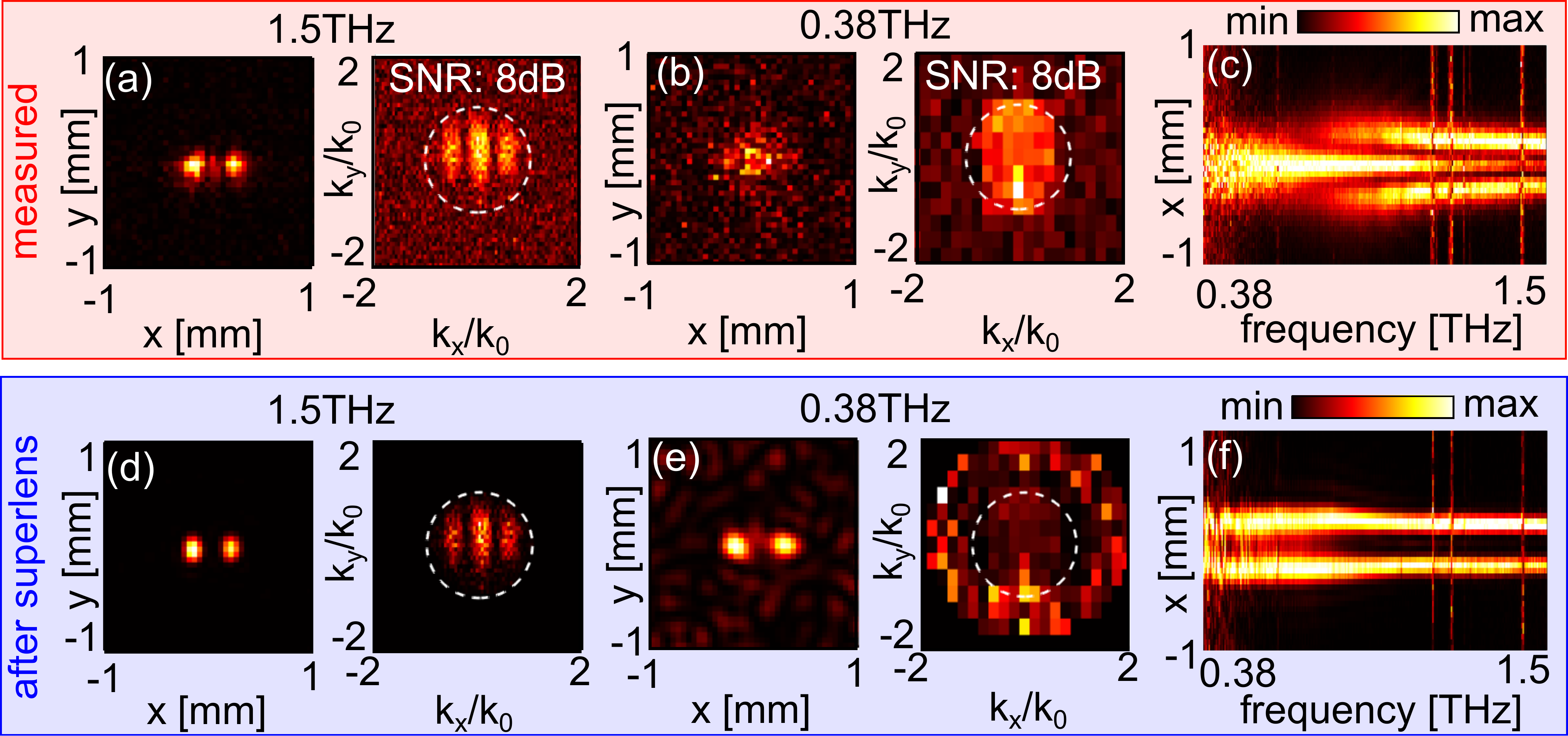}
 \caption{Measured $|E_x|^2$ and $|\tilde{E}_x|$ for two apertures of diameter/separation $d=200\,\mu {\rm m}$, at (a) 1.5\,THz and (b) 0.38\,THz, with ${\rm SNR}$ as labelled. Dashed white circles show $|k| = k_0$. (c) Corresponding intensity profile in $x$ as a function of frequency averaged over $y=0 \pm 100\,\mu{\rm m }$. 
 (d) $|E^{\rm SL}_x|^2$ after the superlens at 1.5\,THz and (e) 0.38\,THz. (g)  Corresponding intensity profile in $x$ as a function of frequency averaged over $y=0 \pm 100\,\mu{\rm m}$.  Vertical artefacts in (c),(f) are  absorption lines due to air humidity.
 }
 \label{fig:fig3}
 \end{figure*}
% %%%%%%%%%%%%%%%%%%%%%%%%%%%%%%%%%%%%%%%%%%%

\subsection*{Experiments}

\begin{figure*}[t!]
\centering
\includegraphics[width=1\textwidth]{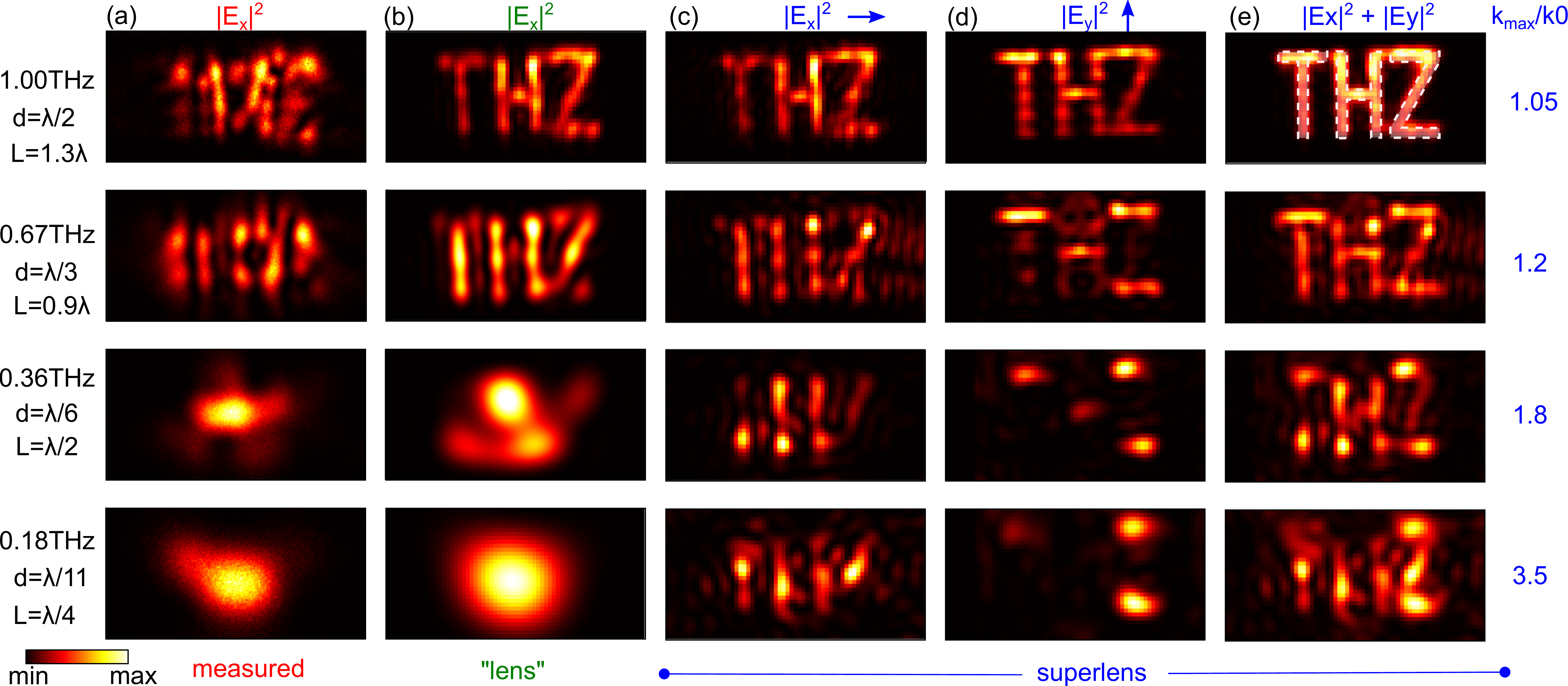}
\caption{Superlens experiment, imaging the letters ``THZ'' (feature size: $d = 150\,{\mu{\rm{m}}}$; detector distance: $L = 440\,{\mu{\rm{m}}}$). (a) Measured $|E_x|^2$ at different frequencies labelled left, with associated  $d$ and $L$ in terms of $\lambda$. (b) Corresponding field after the superlens with $\kmax = k_0$ (lens) and (c) when $\kmax > k_0$ (superlens). (d) Reconstructed images using measured $x-$ and $y-$ polarized fields in (c),(d). The ratio $\kmax/k_0$ for each row is shown on the right. Each window area is 4\,mm $\times$ 2\,mm.}
\label{fig:fig4}
\end{figure*}

Figures~\ref{fig:fig3} and \ref{fig:fig4} showcase our technique on two distinct imaging experiments. Our experiment uses a commercial pulsed THz source (Menlo TERAK15, 0.1-3\,THz). Lenses collimate and focus the THz beam towards a patterned laser-machined samples containing sub-wavelength features ($d=150-200\,\mu{\rm m}$).
The transmitted field amplitude is sampled as a function of the time delay of a probe pulse on a photoconductive antenna which probes the radiating near field. The electric field is polarized in $x$, using the reference frame shown in Fig.~\ref{fig:fig1}. Diffraction limited imaging of the finest feature would require a wavelength of $\lambda/2=d$, \ie,  frequencies of $0.75-1\,{\rm THz}$.
See Supplementary Fig. 2 for detailed images of samples, near field probe, and experimental layout.

Figure~\ref{fig:fig3}(a) and~\ref{fig:fig3}(b) shows the measured intensity $|E_x|^2$ emerging from two apertures (diameter and separation: $200\,\mu{\rm m}$) at a frequency of 1.5\,THz and 0.38\,THz respectively, as well as the associated spatial Fourier transform magnitude $|\tilde{E}_x|$. Figure~\ref{fig:fig3}(c) shows the average measured intensity across $y=0 \pm 100\,\mu{\rm m}$ as a function of $x$ and frequency, and highlights that apertures cannot be  discerned directly, either due to diffraction at higher frequencies, or because of evanescent decay at lower frequencies. 
We calibrate the probe-to-sample distance $L$ by considering the complex field at a frequency above the diffraction limit (here: 1.5 THz), and adjusting $L$ in Eq.~\eqref{eq:SL} to maximize image sharpness (see Supplementary Fig. 3). From $|\tilde{E}_x|$, we then obtain the frequency-dependent SNR via the ratio between the maximum amplitude in the propagating region $|\mathbf{k}|<k_0$, and the average amplitude in the evanescent region $|\mathbf{k}|>k_0$, from which we estimate the experimentally accessible $\kmax$ via Eq.~\eqref{eq:kmax}.  For the two-aperture experiment we find $L=172\,\mu{\rm m}$ (\ie~$L\simeq 0.87\lambda$ at 1.5~THz and $L\simeq 0.22\lambda $ at 0.38~THz) and ${\rm SNR} = 4-14\,{\rm dB}$ between 0.38-1.5\,THz, resulting in $\kmax/k_0= 1- 1.9$ (see Supplementary Fig. 4). 
We then implement the SL via Eq.~\eqref{eq:SL}, followed by spatial low-pass filtering bounded by $\kmax$. Figure~\ref{fig:fig3}(d) and ~\ref{fig:fig3}(e) show $|\Ereconstructedx|^2$ and $|\tilde{E}_x|$ at 1.5\,THz and 0.38\,THz respectively: the  apertures are now resolved. The associated average intensity over the middle of the two apertures (Fig.~\ref{fig:fig3}(f)), shows this procedure works over the entire THz frequency band. 

%%%%%%%%%%%%%%%%%%%%%%%%%%%%%%%%%%%%%%%%%%

Finally, we implement our procedure on a complex, large-area image formed by a laser-written metal sheet containing the letters ``THZ'' (minimum feature size:  $d=150\,\mu{\rm m}$). In this case, the calibration yields $L = 440\,\mu{\rm m}$, SNR=15-25\,dB, and $\kmax/k_0 = 1-3.5$ between 0.2-1\,THz (see Supplementary Fig.~4). Figure~\ref{fig:fig4}(a) shows the measured $|E_x|^2$ at different frequencies as labelled. Figure~\ref{fig:fig4}(b) shows the corresponding retrieved field $|\Ereconstructedx|^2$ through Eq.~\eqref{eq:SL}  at different frequencies with $\kmax=k_0$, \ie, a conventional lens simply reversing phase. At 1.0\,THz (\ie, the diffraction limit), the letters ``THZ'' can be discerned; at 0.67\,THz, only the vertical features are resolved; lower frequencies do not provide a sufficiently sharp image to discern the finer features of the original image, with only a single large spot occurring at 0.18\,THz. Figure~\ref{fig:fig4}(c) shows the corresponding retrieved $|\Ereconstructedx|^2$ at different frequencies with $\kmax/k_0$ as labelled:  vertical features are significantly sharpened. Note that horizontal features do not let $E_x$ through: the slits forming the letters act as parallel plate waveguides with width smaller than half a wavelength, in which solely the TEM mode polarized perpendicularly to the thinnest features can propagate, so that thin features in $y$ ($x$) only appear for the $E_x$ ($E_y$) field. As a result, for $x-$polarized fields the letters' vertical ($y$-oriented) features are clearest, in agreement with simulations (see Supplementary Fig.~5). We repeat the above procedure for a polarization oriented in $y$ relative to the sample orientation, and plot the corresponding $|E_y|^2$ in Fig.~\ref{fig:fig4}(c) to resolve the horizontal features of each letter. The full image is obtained by summing the two contributions: Fig.~\ref{fig:fig4}(d) show the resulting $|E_x|^2 + |E_y|^2$ distribution, clearly showing the emergence of the letters ``THZ'' at all frequencies, down to $\lambda/7$. These results are in agreement with simulations of the transmitted subwavelength pattern (see Supplementary Fig.~5).
Remarkably, the resolution achieved in Fig.~\ref{fig:fig4} is higher than that in Fig.~\ref{fig:fig3}, even though the near field antenna's distance to the object was more than doubled, thanks to a higher  SNR exceeding the loss from increased evanescent decay.

\section*{Discussion}

In this paper, we have presented a novel superlensing approach which numerically amplifies measured evanescent fields to obtain complex images with subwavelength features, limited only by the noise of the instrument. We presented experiments illustrating its implementation at THz frequencies using commercially available facilities. Our approach can be adapted to suit any near field experiment which measures amplitude and phase, immediately providing a pathway for increasing the imaging resolution of near field setups at any frequency. High resolution near field measurements in the radiating rather than reactive near field will allow accurate near-field imaging without perturbing the intrinsic field of structures strongly susceptible to local disturbances, such as high-Q/topological resonators~\cite{van2021unveiling, yang2022topology} and photonic crystal defects~\cite{akiki2020high}.

\section*{Methods}

\subsection*{Experimental setup} 

 A schematic of the experimental setup is shown in Supplementary Fig. 2. We use a commercially available THz-TDS System (Menlo TERAK15), which relies on THz emission from biased photoconductive antennas that are pumped by fiber-coupled near-infrared pulses (red line; pulse width: 90\,fs; wavelength: 1560\,nm).  Terahertz lenses collimate and focus the beam towards the sample. The THz field emerging from the THzLC is sampled as a function of the time delay of a fiber-coupled probe pulse on another photoconductive antenna, which forms the THz detector. The electric field is polarized in $x$, using the sample orientation and reference frame shown in Fig. 1. A moveable, fiber-coupled near-field (NF) detector module enables the measurement of the $x$-polarized electric field at the output of the laser-machined samples. The near-field is spatio-temporally resolved at every point via a raster scan (step size: 25-50\,{$\mu{\rm m}$}). Fast Fourier transforms of the temporal response at each pixel position provide the spectral information (spectral resolution: 5-8\,GHz). 
\subsection*{Data Availability} 
The data that support the findings of this study are available from the corresponding author upon reasonable request.

\bibliography{main_ARXIV}

\section*{End Notes}

\subsection*{Acknowledgments}
This work is funded in part by the Australian Research Council Discovery Early Career Researcher Award (DE200101041). The authors thank Angus Michael O’Grady and Gleb Kozlov for fruitful discussions. The authors thank Benjamin Johnston from the Optofab Node of the Australian National Fabrication Facility for fabricating the laser machined samples. 

\subsection*{Author Contributions}
A.T. and B.T.K. conceived the idea. A.T. performed the experiments and simulations. B.T.K. derived the noise dependent maximum spatial frequency resolution limit. A.T. and B.T.K. wrote the manuscript. A.T. directed the project. 

\subsection*{Competing Interests}
The authors declare no competing financial interests. 

\subsection*{Materials \& Correspondence}
Correspondence to Alessandro Tuniz.

\clearpage

\newpage 

\clearpage

\onecolumngrid

\renewcommand{\figurename}{Supplementary Fig.\,\kern-3pt}

\setcounter{figure}{0}

\vspace{-1.1cm}

\section{Supplementary Text}

\subsection*{Noise limit derivation}
Defining  $k_{\perp}^2 = k_x^2+k_y^2$ so that $k_{z}  =  \sqrt{k_0^2 - k_{\perp}^2}$, let $\tilde{T}$ be the transfer function of the fields along $z$:
\begin{equation}
\tilde{T}(k_x,k_y,z) =  {\rm exp}(i k_zz)
\end{equation} such that $\tilde{E}(k_x,k_y,z)=\tilde{T}(k_x,k_y,z)\tilde{E}(k_x,k_y,z=0)$ (which is propagation in free space) and conversely $\tilde{E}(k_x,k_y,z=0)=\tilde{T}(k_x,k_y,-z)\tilde{E}(k_x,k_y,z)$ (which is the superlensing procedure). 
 For propagating fields $k_{\perp}<k_0$ and
\begin{equation}
|T(k_{\perp}<k_0)| = 1.
\end{equation}
For evanescent fields, $k_{\perp}>k_0$ and
\begin{equation}
|T(k_{\perp}>k_0)| = {\rm exp}(\pm|k_z| z),
\end{equation}
with a $-$ sign for the evanescent decay (transfer from object to measurement distance) and $+$ sign for the superlensing procedure.

At a distance $z$ from the source, the spatial Fourier transform of the measured field is given by

\begin{equation}
\tilde{E}_M(k_x,k_y) = \tilde{E}_m(k_x,k_y) + \tilde{\delta}_m(k_x,k_y)
\end{equation}
where $\tilde{E}_m(k_x,k_y)$ is the actual field (\ie, in the absence of noise), and $\tilde{\delta}_m(k_x,k_y)$ is the noise due to the measurement instrument, where typically $|\tilde{\delta}_m(k_x,k_y)| \ll {\rm{max}}|\tilde{E}_m(k_x,k_y)|$.
The reconstructed field is given by 
\begin{equation}
    \tilde{E}^{\rm SL}(k_x,k_y) = T(k_x,k_y,-z)\tilde{E}_m(k_x,k_y) + \tilde{T}(k_x,k_y,-z)\tilde{\delta}_r(k_x,k_y)
\end{equation}
The first product is exactly the object field we wish to obtain, while the second term is the noise in the reconstructed field  $\tilde{\delta}^{\rm SL}=\tilde{\delta}_m\exp(|k_zz|
)$,
where we omit the explicit $(k_x,k_y)$ dependence for brevity. Compared to the measurement noise, the (amplitude) noise penalty in dB of the procedure is thus
\begin{equation}
    \Delta_{\rm noise}=10\log_{10}\left|\frac{\tilde{\delta}^{\rm SL}}{\tilde{\delta}_m}\right|=10\log_{10}(\exp(k_zz))=\frac{10k_zz}{\log(10)}.
\end{equation}
When the noise penalty exceeds the initial signal to noise ratio of the measurement
SNR, the image becomes dominated by noise. This occurs when
\begin{equation}
    {\rm SNR}<\Delta_{\rm noise}=\frac{10z\sqrt{k_\perp^2-k_0^2}}{\log(10)}
    \label{eq:snrnoise}
\end{equation}
The right hand term increases steadily with $k_\perp$, and the limiting value of $k_\perp=\kmax$ at which equality is achieved can be obtained by rearranging Eq.~\eqref{eq:snrnoise} as 
\begin{equation}
\kmax  =  k_0 \sqrt{1+ \left(\frac{\lambda}{z}\frac{\log 10}{20 \pi}{\rm SNR}\right)^2},
\label{eq:kmaxsupp}
\end{equation}
which corresponds to Eq.~(4) in the main manuscript. Filtering out spatial frequencies higher than $\kmax$ ensures that the noise penalty due to the superlensing procedure does not generate noise levels above the measurement's signal level, and thus ensures clear images.

\subsection*{Noise limit examples}

\begin{figure}[h!]
\centering
\includegraphics[width=1\textwidth]{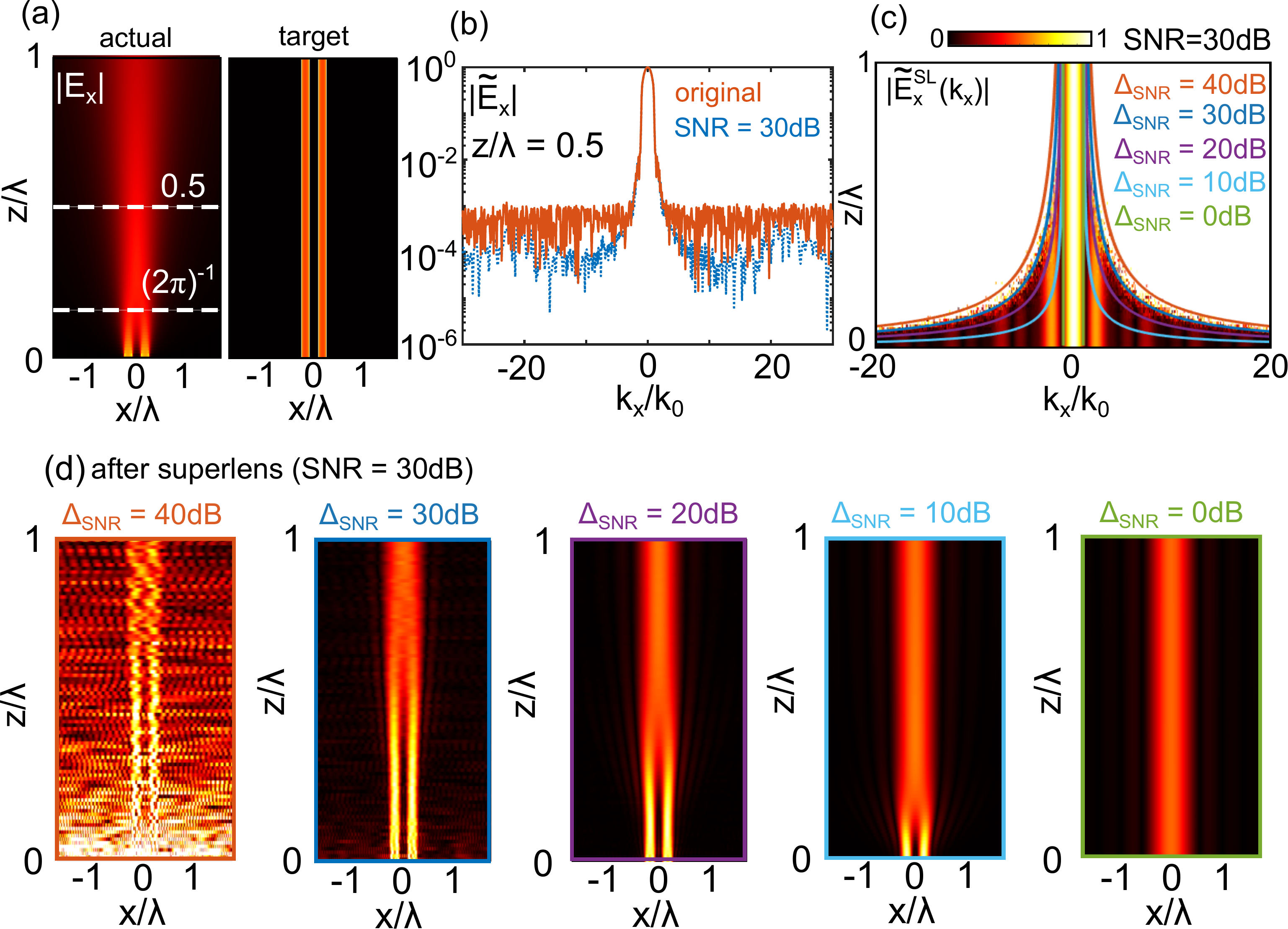}
\caption{Numerical example illustrating the effect of the SNR, when applying the superlens procedure of Eq.~(3) in the manuscript. (a) Left: raw simulation of the  amplitude $E_x(x)$ as a function of $z/\lambda$ for the double aperture case of Fig.~2 in the manuscript, before any superlens procedure. Right shows the target amplitude at the source, that is, what a figure should look like after perfect, noiseless superlensing: both apertures are resolved for all choice of $z/\lambda$. (b) Dashed blue line shows an example $|\tilde{E}_x|$ at a distance $z/\lambda=0.5$, as per Fig. 2(b) of the manuscript. The red line shows the same field after adding random amplitude and phase resulting in a flat SNR of 30\,dB. (c) Calculated normalized amplified spatial Fourier transform for the flat SNR = 30\,dB, as a function of its normalized propagation length $z/\lambda$. Color scale has been saturated to 1 for clarity. Solid curves show Eq.~\eqref{eq:delta_SNR} choosing different values of $\Delta_{\rm SNR}$ as labelled. Note that the case $\Delta_{\rm SNR} = {\rm SNR} = 30\,{\rm dB}$ corresponds to the boundary between where high spatial frequencies have a comparable magnitude to propagating waves.  (d) Resulting $E_x^{SL}(x)$ after applying the superlens procedure, using a low-pass filter function bounded by $k_x/k_0$ as per Eq.~\ref{eq:delta_SNR} for different values of $\Delta_{\rm SNR}$ as labelled. If $\Delta_{\rm SNR}>{\rm SNR}$, the aperture images are plagued by noise; if $\Delta_{\rm SNR}\leq {\rm SNR}$, the maximum distance at which the retrieval procedure produces the image is gradually reduced. The maximum useful $z/\lambda$ occurs at the point in which  $\kmax/k_0$ contains the feature needed (here, $\kmax/k_0 = 2.5$, which for $\Delta_{\rm SNR}=30\,{\rm dB}$ occurs at $z/\lambda=0.5$.)}
\label{fig:noise_effects}
\end{figure}

We now consider an example to showcase the implications of  Eq.~\eqref{eq:kmaxsupp} by expanding our analysis on the simulations associated with Fig.~2 of the manuscript. Supplementary Fig.~\ref{fig:noise_effects}(a) (left) shows a simulation of the amplitude $E_x(x)$ as a function of normalized propagation length $z/\lambda$ for the double aperture case of Fig.~2 in the manuscript, before any superlens procedure. Note in particular that inside the interval $(2\pi)^{-1}<z/\lambda<1$, \ie, in the radiating near field, the two apertures are not resolved. On the right of Supplementary Fig.~\ref{fig:noise_effects}(a) we show the target amplitude at the source, where both apertures are resolved for all choice of $z/\lambda$.  The blue line in Supplementary Fig.~\ref{fig:noise_effects}(b) shows the the Fourier transform $|\tilde{E}_x|$ at a distance $z/L=0.5$, as per Fig. 2(b) of the manuscript. To enable a clear analysis, we add random amplitude and phase, resulting in a nominally flat SNR of 30\,dB, shown as a line curve in Supplementary Fig.~\ref{fig:noise_effects}(b). We then perform the superlensing procedure without any filtering, for different propagation lengths $z$, always adding white noise such that SNR=30\,dB before amplification. The resulting $|\tilde{E}^{\rm SL}_x(k_x,k_y)|$, normalized to $|\tilde{E}^{\rm SL}_x(0,0)|$, is shown in Supplementary Fig.~\ref{fig:noise_effects}(c), where the color scale is saturated at unity. For short lengths $z/\lambda$, the procedure yields the required high spatial frequency components. For increasingly long propagation lengths however, amplified evanescent wave amplitudes are greater than those of the propagating waves (white regions). Equation~\ref{eq:kmaxsupp} predicts the boundary between these two regions for SNR=30\,dB. To show this,  Supplementary Fig.~\ref{fig:noise_effects}(c) also shows contour lines of the function
\begin{equation}
k_x/k_0 = \sqrt{1+ \left(\frac{\lambda}{z}\frac{\log 10}{20 \pi}{\rm \Delta_{\rm SNR}}\right)^2},
\label{eq:delta_SNR}
\end{equation}
for different choices of $\Delta_{\rm SNR}$ as labelled.  For $\Delta_{\rm SNR} = {\rm SNR} = 30\,{\rm dB}$ (dark blue), where the $\Delta_{\rm SNR}$ value used in Eq.~\eqref{eq:delta_SNR} is that of the actual SNR, Eq.~\eqref{eq:delta_SNR} indeed yields the spatial frequency boundary inside which the evanescent field amplitude after the virtual SL remain below that of propagating waves. Decreasing $\Delta_{\rm SNR}$ narrows the available range of $\kmax$, as expected, with $\Delta_{\rm SNR}=0\,{\rm dB}$ corresponding to $\kmax = k_0$. Given a set of experimental conditions, Eq.~\eqref{eq:kmaxsupp} can thus be used as a first estimate of the low-pass spatial frequency boundary to include after the superlensing procedure, and which can be fine-tuned until an suitable image is obtained.  This procedure contains a trade-off between image sharpness and noise: A wider $\kmax/k_0$ boundary has the effect of deteriorating the image by using amplified noisy spatial frequencies, as shown in Supplementary Fig.~\ref{fig:noise_effects}(d), left; a narrower $\kmax/k_0$ boundary removes high spatial frequencies needed to resolve fine feature sizes, as shown in Supplementary Fig.~\ref{fig:noise_effects}(d), right.

\newpage

\section{Supplementary Figures}

%%%%%%%%%%%%%%%%%%%%%%%%%%%%%%%%%%%%%%%%%%

%%%%%%%%%%%%%%%%%%%%%%%%%%%%%%%%%%%%%%%%%%
\begin{figure}[h!]
\centering
\includegraphics[width=0.6\textwidth]{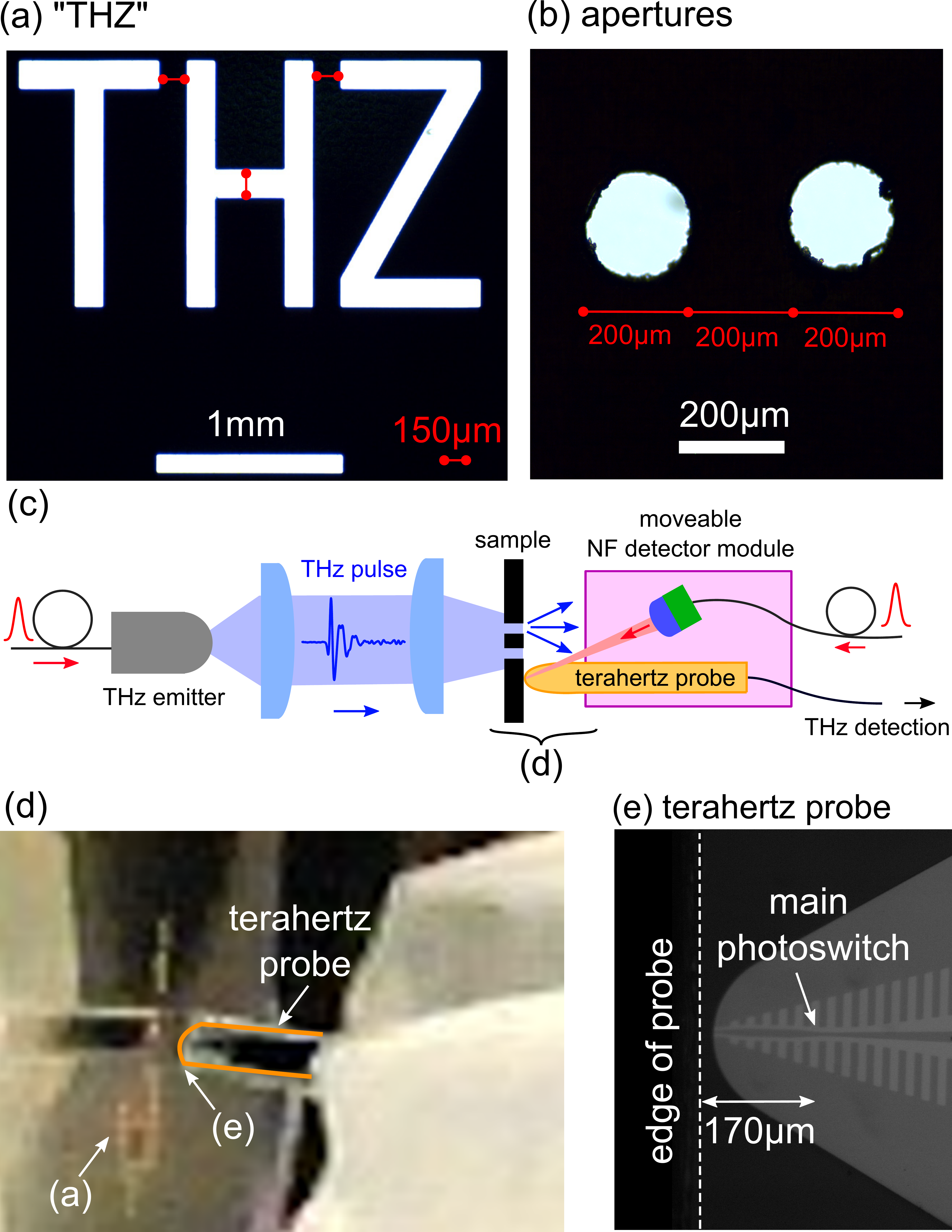}
\caption{(a) Optical microscope images of (a) the laser-machined ``THZ'' letters used in Fig. 4 of the manuscript, and (b) the two apertures used for the experiments in Fig. 3 of the manuscript, highlighting the minimum feature sizes in each case. Black: metal; white: air gaps. (c) Schematic of the experimental setup.  We use a commercially available THz-TDS System (Menlo TERAK15), which relies on THz emission from biased photoconductive antennas that are pumped by fiber-coupled near-infrared pulses (red line; pulse width: 90\,fs; wavelength: 1560\,nm).  Terahertz lenses collimate and focus the beam towards the sample. The THz field emerging from the THzLC is sampled as a function of the time delay of a fiber-coupled probe pulse on another photoconductive antenna, which forms the THz detector. The electric field is polarized in $x$, using the sample orientation and reference frame shown in Fig. 1. A moveable, fiber-coupled near-field (NF) detector module enables the measurement of the $x$-polarized electric field at the output of the laser-machined samples. (d) Photograph of the near-field terahertz probe  as it scans the surface of the sample. (e) Microscope image of the near-field terahertz probe, showing the location of its main photoswitch, which is at a nominal distance of $170\,\mu {\rm m}$ from the edge of the probe -- this also means for this probe used at right angle, only object-to-photoswitch distances $L>170\mu$m are achievable. Note that small variations in the alignment of the pulses used in the near field detector module -- performed anew before starting each scan -- can affect both $L$ and the SNR.}
\end{figure}
%%%%%%%%%%%%%%%%%%%%%%%%%%%%%%%%%%%%%%%%%%

%%%%%%%%%%%%%%%%%%%%%%%%%%%%%%%%%%%%%%%%%%
\begin{figure}[h!]
\centering
\includegraphics[width=0.55\textwidth]{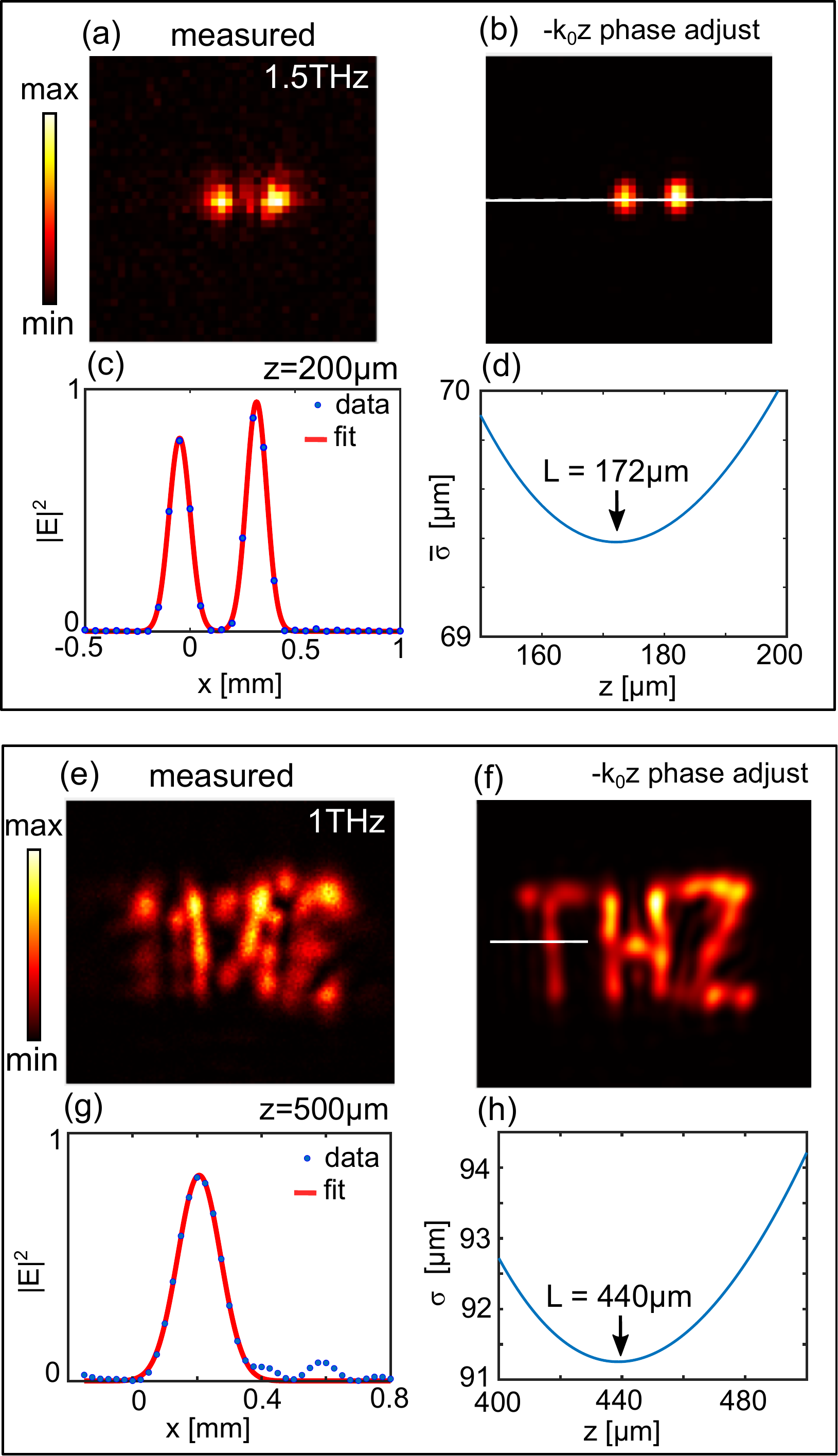}
\caption{Retrieval of the sample-to-switch distance $z=L$. (a) Measured $|E_x|$ at 1.5\,THz for the two-aperture experiment, where the spatial frequencies which resolve the aperture features are propagating. (b) Example image when the phase of $|\tilde{E}_x|$ is adjusted by $-k_0z$, where $z=200\,\mu{\rm m}$. We consider its field magnitude through the apertures' center (white dotted line), and show it as blue circles in (c). Red curves in (c) show a Gaussian fit to a double Gaussian function, used to obtain the average standard deviation $\overline{\sigma}$. (d) Average fitted  $\overline{\sigma}$ as a function of $z$. (e)-(h) Same as (a)-(d), performed for the ``THZ'' sample. Here we use a single feature, highlighted by the white line in (f), and fit a single gaussian. In this case, $z=L=440\,\mu{\rm m}$. In both cases, this numerical processing of our experimental data functionally corresponds to adjusting the focus of a conventional lens, until the sharpest image is obtained.}
\end{figure}
%%%%%%%%%%%%%%%%%%%%%%%%%%%%%%%%%%%%%%%%%%

\newpage

%%%%%%%%%%%%%%%%%%%%%%%%%%%%%%%%%%%%%%%%%%
\begin{figure}[h!]
\centering
\includegraphics[width=1\textwidth]{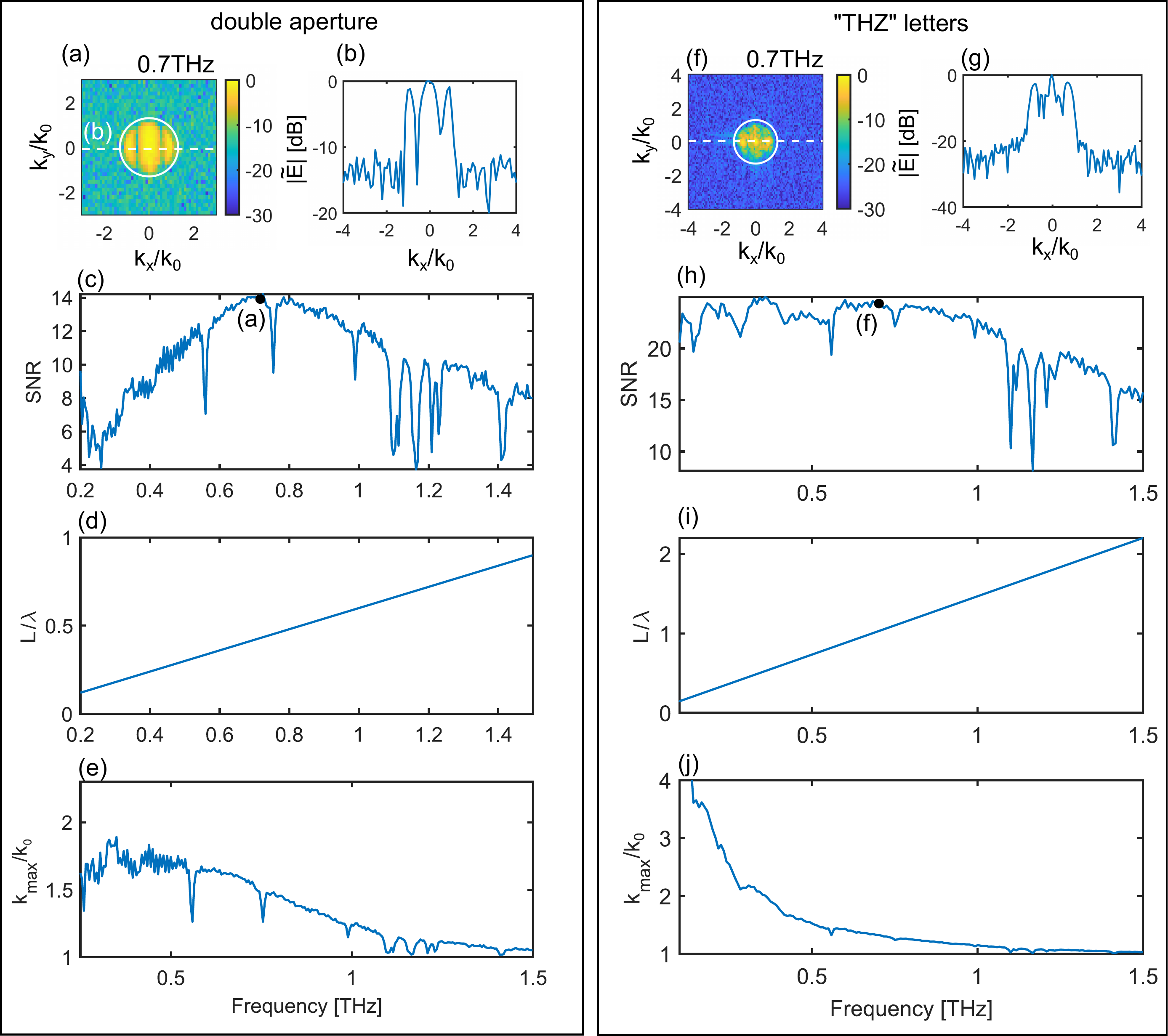}
\caption{Overview of the method for obtaining the experimental $\kmax/k_0$ as a function of frequency for the double aperture experiment of Fig.~4 of the manuscript. (a) Measured $|\tilde{E}_x|$ as a function of $k_x/k_0$ and $k_y/k_0$ at an example frequency 0.7\,THz. (b) $|\tilde{E}_x|$ as a function of $k_x$ for $k_y=0$ (white dashed line in (a)), where SNR = 14\,dB. (c) Extracted ${\rm SNR} = {\rm max}(|\tilde{E_x}|)/\langle|\tilde{E}_x|(|\mathbf{k}|>k_0)\rangle$ as a function of frequency. The black circle shows the value obtained from (a). The boundary between the propagating and evanescent regions, used in the SNR measurement, is shown as a white circle in (a). Note the maximum SNR of 14\,dB at 0.7\,THz. (d) Associated $L/\lambda$ as a function of frequency. (e) Resulting $\kmax/k_0$ as a function of frequency, calculated using Eq.~(4) of the manuscript. Here $\kmax/k_0=$1--1.8 between 0.2--1.5\,THz.  (f)--(j) Same as (a)--(e), applied to the ``THZ'' letters experiment shown in Fig.~5 of the manuscript. Note that, although the sample-to-detector distance $L$ is about twice as long here compared to the double-aperture experiment (implying that high spatial frequencies have decayed more), the SNR reaches values that are nominally 10\,dB higher so that higher spatial frequencies can be amplified ($\kmax/k_0 > 3$), as per Eq.~(4) of the manuscript.}
\end{figure}
%%%%%%%%%%%%%%%%%%%%%%%%%%%%%%%%%%%%%%%%%%

%%%%%%%%%%%%%%%%%%%%%%%%%%%%%%%%%%%%%%%%%%
\begin{figure}[h!]
\centering
\includegraphics[width=\textwidth]{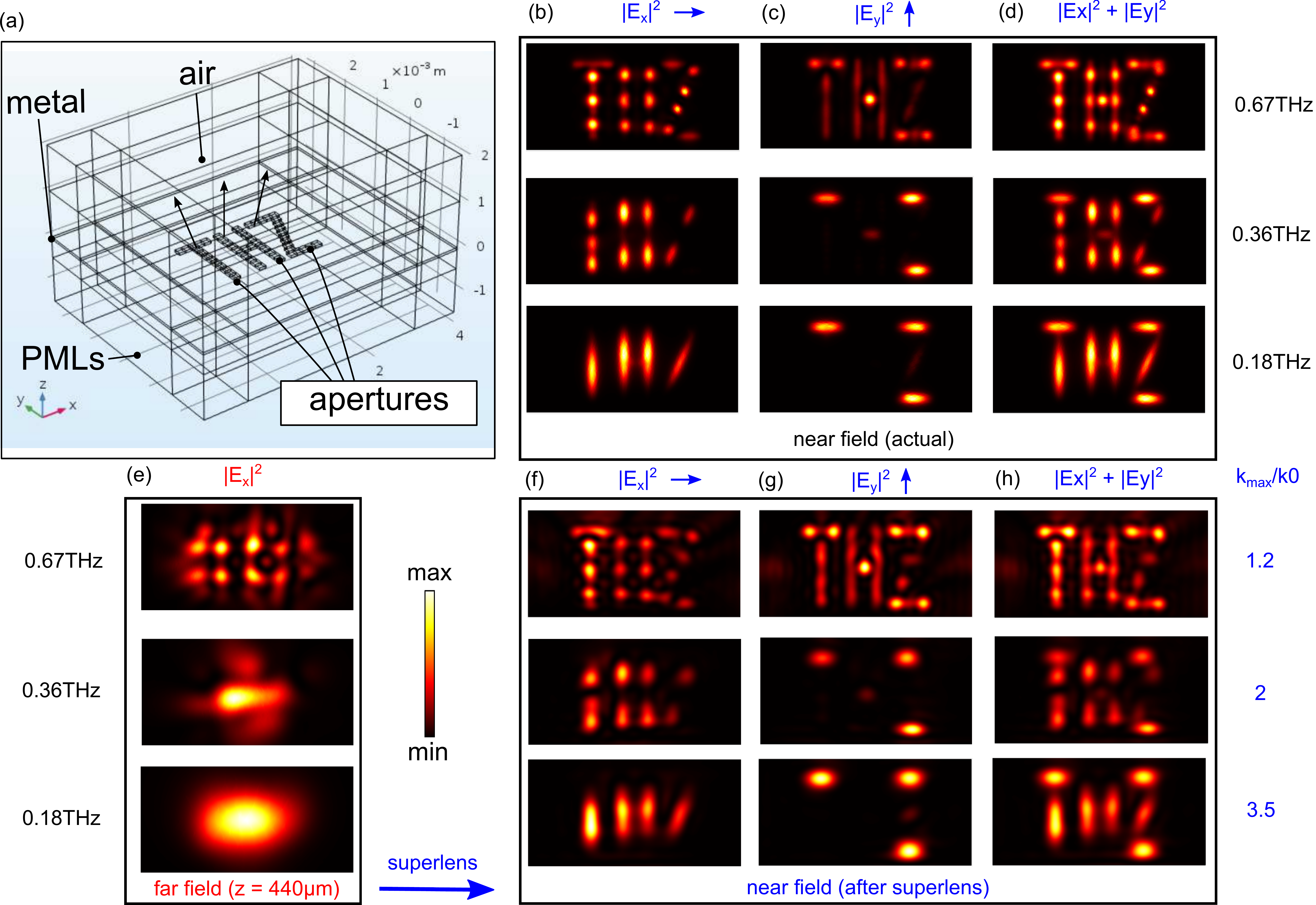}
\caption{3D finite element simulations (COMSOL) modelling the field emerging from the letters ``THZ'' of Fig. 4 in the manuscript, and comparing the superlens procedure result with the near field directly taken from the simulations. (a) Our model replicates the sample used, and considers the electric field emerging from apertures within a metal sheet (metal thickness: $50\,\mu {\rm m}$; aperture width: $150\,\mu{\rm m}$) suspended in air. Perfectly matched layers (PMLs) suppress reflections at the boundaries. (b) Simulated $|E_x|^2$, (c) $|E_y|^2$, and (d) their sum $|E_x|^2+|E_y|^2$ at a distance $z=50\,\mu{\rm m}$ from the metal sheet. (e) Simulated $|E_x|^2$ at a distance $z=440\,\mu{\rm m}$ from the metal sheet, for the frequencies as labelled. Also shown are the images obtained when applying the superlens procedure to the numerically calculated fields at $z=440\,\mu{\rm m}$, showing the resulting (f) $|E_x|^2$ (d) $|E_y|^2$, and (e) $|E_x|^2+|E_y|^2$, with $\kmax/k_0$ as labelled, for comparison with our experiment. Note that the directly simulated near-fields are comparable to those obtained after applying the superlens procedure in the radiating near field, and display many of the subtle features observed in our experiment, \cf Fig. 4 of the manuscript. Each window area is 4\,mm $\times$ 2\,mm. }
\end{figure}
%%%%%%%%%%%%%%%%%%%%%%%%%%%%%%%%%%%%%%%%%%

\end{document}